\documentclass{iau}

\usepackage{amsmath}
\usepackage{graphicx}
\usepackage{multirow}
\usepackage{comment}
\usepackage{subcaption}
\usepackage{aas_macros}

\begin{document}

\lefttitle{Puja Majee}
\righttitle{A complex type-II radio burst observed with the MWA} 
\jnlPage{1}{7}
\jnlDoiYr{2024}
\doival{10.1017/xxxxx}
\volno{388}
\pubYr{2024}
\journaltitle{Solar and Stellar Coronal Mass Ejections}

\aopheadtitle{Proceedings of the IAU Symposium}
\editors{N. Gopalswamy,  O. Malandraki, A. Vidotto \&  W. Manchester, eds.}

\title{On Using Gradient Dynamic Spectra (GraDS) to Study Type-II Solar Radio Bursts}

\author{Puja Majee$^1$, Devojyoti Kansabanik$^{2,3}$, Divya Oberoi$^1$}
\affiliation{$^1$National Centre for Radio Astrophysics, Tata Institute of Fundamental Research, S. P. Pune University Campus, Pune, India, 411007}
\affiliation{$^2$Cooperative Programs for the Advancement of Earth System Science, University Corporation for Atmospheric Research, Boulder, CO, USA}
\affiliation{$^3$NASA Jack Eddy fellow hosted at the Johns Hopkins University Applied Physics Laboratory, 11001 Johns Hopkins Rd, Laurel, MD, USA}

\begin{abstract}
Solar type-II radio bursts are coherent plasma emissions arising from magnetohydrodynamic shocks produced by either coronal mass ejections (CMEs) or flares. Type-II bursts sometimes show split-band emissions in the dynamic spectrum. When these split-band emissions come from regions just upstream and downstream of the shock, type-II band-splitting can be used as an important tool for estimating magnetic fields at the shock front. Earlier studies have shown that only $\sim$20\% of the type-IIs show morphologically similar split-bands. Imaging studies can unambiguously identify such instances, though they remain very rare. Here we suggest a useful approach to augment dynamic spectra based studies by also examining the Gradient Dynamic Spectra (GraDS) of type-II emission. We also verified the conclusions of this approach against those from an imaging study.

\end{abstract}

\begin{keywords}
Coronal Mass Ejections, Shocks, Solar Radio Burst, Radio Imaging
\end{keywords}

\maketitle

\section{Introduction}
Type-II solar radio bursts are plasma emissions generated by magnetohydrodynamic (MHD) shocks driven by energetic eruptions from the Sun, such as flares and coronal mass ejections (CMEs). The MHD shocks accelerate electrons to high speeds which excite  Langmuir waves, eventually resulting in plasma emissions at the fundamental (F) and harmonic (H) of the local plasma frequency \citep{McLean1985}. Type-II emissions have been observed from metric (m-), decametric (Dm-) to kilometric (km-) wavelengths arising over a wide range of coronal heights and interplanetary medium. Association with shocks of large-scale solar eruptions makes studying type-IIs extremely attractive from the perspective of understanding space weather, particularly for determining solar energetic particle (SEP) events. Several studies have concluded that metric type-IIs mostly originate from CME-driven shocks \citep[e.g.,][]{Gopalswami2006, Kumar2023} and hence can be used to understand the energetics, kinematics, and dynamics of the shocks produced by the CMEs.

Type-II bursts often appear as two (F and H) slowly drifting bands (with a typical drift rate $\lesssim 1$MH/s) in the dynamic spectrum. Each of these bands is often seen to be split into two or multiple parallel lanes, known as band-splitting and multi-lane emissions, respectively. Type-IIs can also show short-lived, fragmented fine structures resulting from the interaction between the shock and the inhomogeneities in the coronal density structures \citep{Jasmina2020}. Despite being known for several decades, the origin of the band-splitting is still under debate. Several interpretations have been presented in this regard. One of the leading hypotheses, proposed by \citet{Smerd1974}, is the so-called ``upstream/downstream scenario". According to this interpretation, band-splitting occurs when shock-accelerated electrons travel both ahead and behind the shock front. They produce plasma emission at slightly different frequencies due to different electron densities in the upstream (ahead) and downstream (behind) regions of the shock front. Another interpretation is that the emissions originate from different parts of the shock that propagate through the coronal medium with different densities \citep{mclean1967}. An attractive aspect of the upstream/downstream scenario is that it allows one to determine the shock properties (i.e. shock Mach number, density jump), ambient plasma properties (i.e. Alfv\'{e}n velocity, local electron density), magnetic field strength even in the upper corona and inner heliosphere where few remote observational probes to estimate the physical properties of this magnetized plasma medium are available \citep{Smerd1974,Smerd1975}. 

In a study of coronal and upper coronal (UC)/interplanetary (IP) type-II bursts, \citet{Vrsnak2001, Vrsnak2002, Vrsnak2004} have estimated the shock properties, Alfv\'{e}n speed, magnetic field strength, and their evolution with coronal height using the band-splitting of type-II emissions under the assumption of upstream/downstream scenario. To avoid the possibility of the coincidence of two different emission lanes originating from different parts of the large extended shock front, the authors laid down some detailed criteria for a type-II emission to be regarded as band-splitting. These criteria are -- 1. synchronized temporal and spectral evolution; 2. similar intensity variation 3. frequency ratio differs from 2. They imply a physical scenario where the two lanes originate from two closely spaced radio sources moving in tandem through coronal regions with similar conditions, except for slightly different electron densities. In a sample of 112 metric type-IIs and 139 of UC/IP type-IIs, the authors reported 20\% of metric type-IIs and 17\% of UC/IP type-IIs showing band-splitting emissions satisfying these criteria. 

Many studies have used the band-splitting of type-II bursts to estimate the properties of shocks, Alfv\'{e}n waves, and magnetic field strengths and try to understand the conditions for the occurrence of type-II bursts \citep[e.g.,][etc.]{Cho2011,Ramesh2022,Ndacyayisenga2023}. However, establishing that a given type-II does indeed have band-splitting based on the detailed criteria given by \cite{Vrsnak2001} has been done in a few studies, potentially leading to incorrect conclusions. While the dynamic spectra (DS) cannot provide any spatial information about the emission, a careful examination of the emission lanes can indeed help in exploring the possibility of the radio sources being located in the regions upstream and downstream of the shock.

Instead of visual identification of morphological similarities based on the dynamic spectrum, which can have significant human bias, we tried to develop a systematic approach of identifying the upstream/downstream scenario from the dynamic spectrum to reduce this human bias. Section \ref{sec:method} describes the observation and proposed method. Section \ref{sec:res_&_dis} presents the results and discussion followed by the conclusion from this work in Section \ref{sec:conclusion}.

\section{Observation and Methodology for Identification of Band-splitting Scenario}\label{sec:method}
\begin{figure*}
    \centering
    \includegraphics[width = 0.475\textwidth]{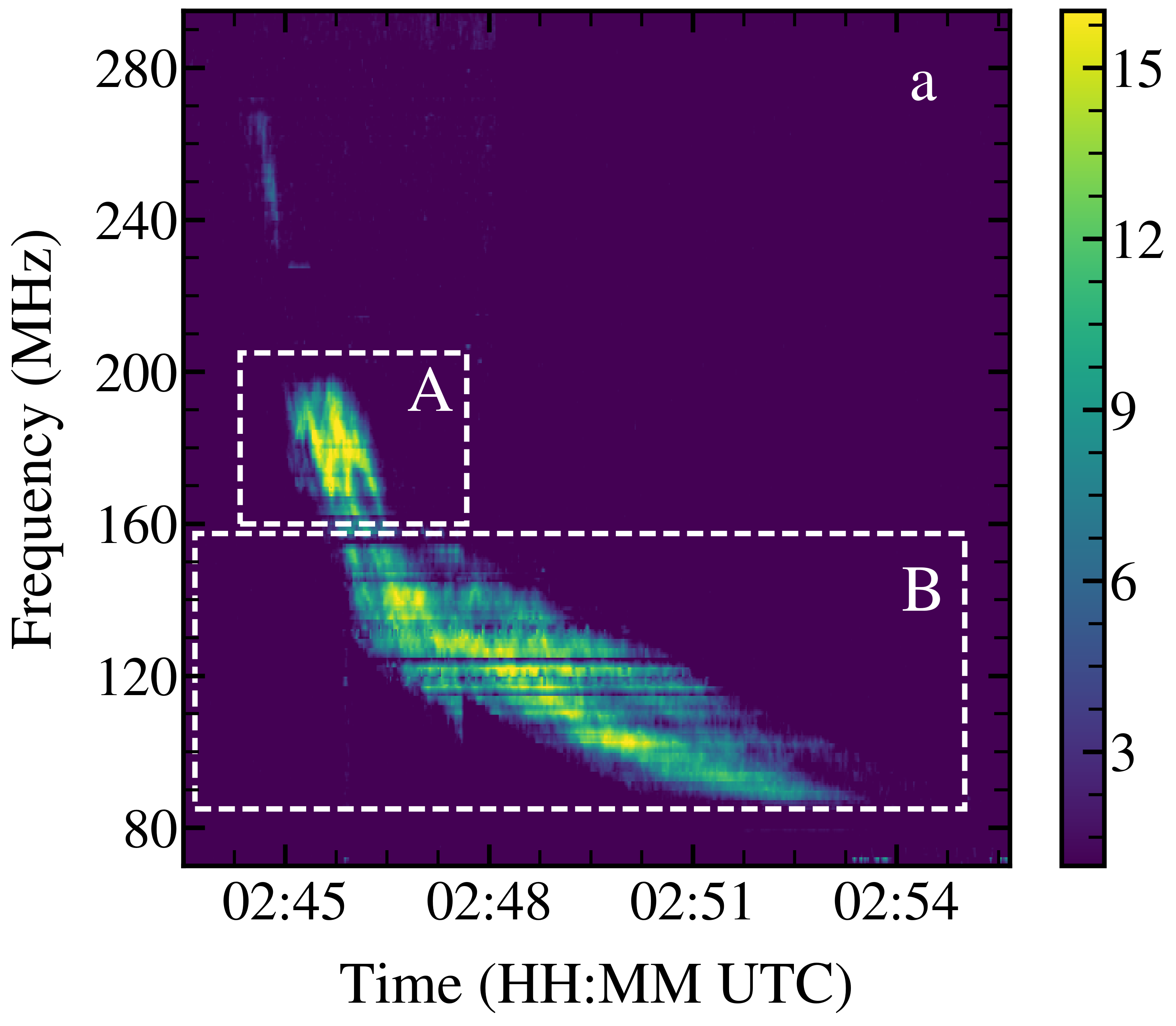}
    \includegraphics[width = 0.5\textwidth]{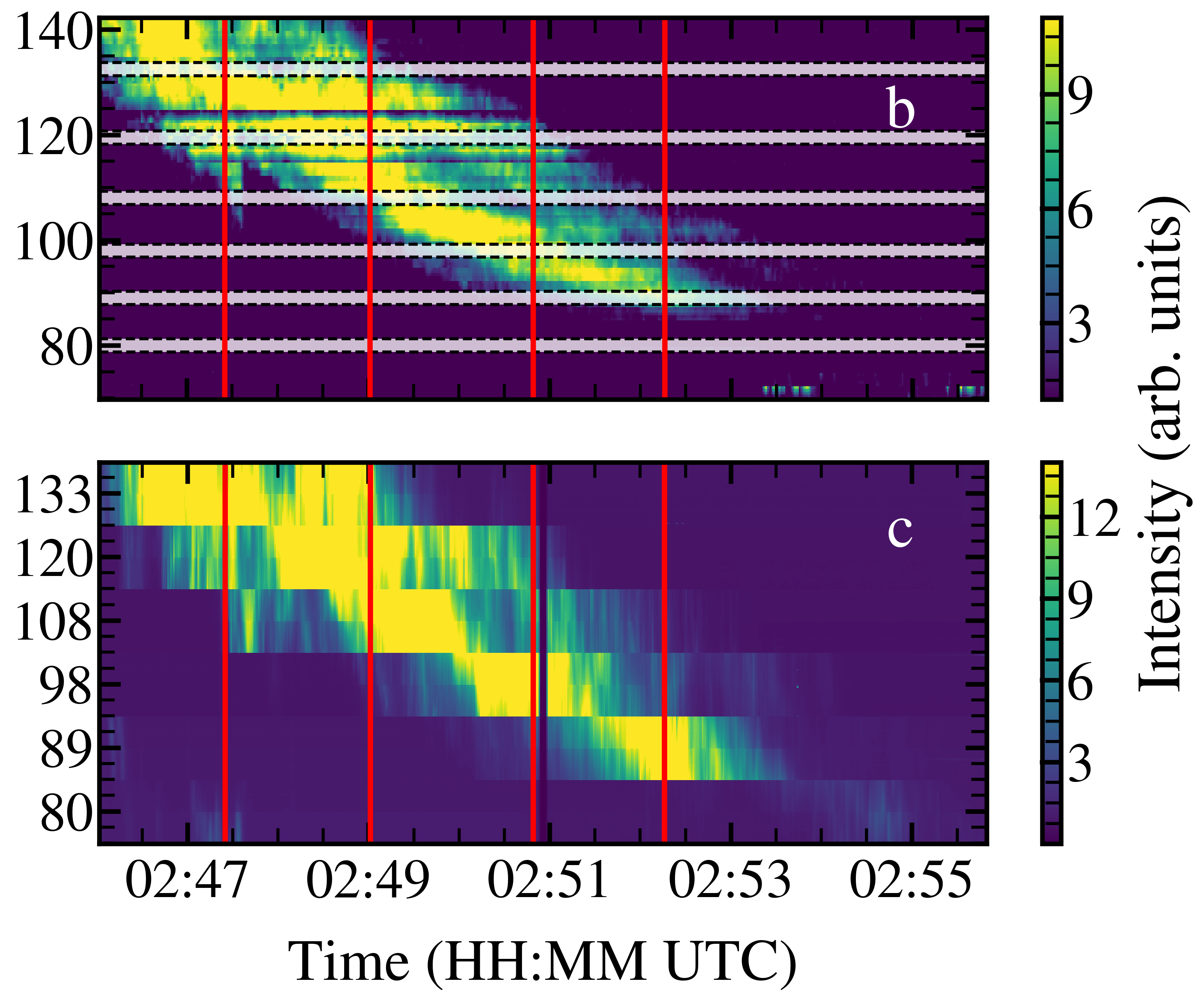}
    \caption{Left panel (a): The dynamic spectrum of the type-II harmonic emission captured by HiRAS-2 spectrograph. The two white dashed boxes (A and B) show the parts of the emission that have been examined in detail to distinguish emission lanes using gradient analysis. Top right panel (b): The lower part of the HiRAS-2 dynamic spectrum overlaps with the MWA observations in six spectral bands highlighted with white rectangles. Bottom right panel (c): The MWA DS in the six spectral bands. The red solid lines in the right panels indicate the four timestamps corresponding to four panels in Figure \ref{fig:aia_mwa}.      
    \label{fig:hiras_mwa_ds}}
\end{figure*}

Here, we present a type-II solar radio burst from 28 September 2014 observed with the Murchison Widefield Array \citep[MWA;][]{Tingay2013} covering frequency range 80 -- 133 MHz with a temporal and spectral resolution of 0.5 seconds and 40 kHz, respectively. A detailed spectroscopic snapshot imaging study of this type-II has been presented in \citet{Bhunia_2023}. To cover a larger spectral span, we use dynamic spectrum from the Hiraiso Radio Spectrograph (HiRAS) \citep{Kondo1995}, which monitors the Sun over the wide frequency range from 25--2500 MHz. HiRAS comprises three independent antenna systems covering different spectral ranges (HiRAS-1: 25-70 MHz; HiRAS-2: 50-550 MHz; HiRAS-3: 500-2500 MHz). Here, we have used HiRAS-2 which provides a temporal and spectral resolution of 1 second and 0.5 MHz, respectively. Figure \ref{fig:hiras_mwa_ds}(a) shows the HiRAS dynamic spectrum of the harmonic emission. To make the emission lanes stand out visually, boxcar smoothing has been done over 5 seconds and 2 MHz. Figure \ref{fig:hiras_mwa_ds}(b) and (c) zoom in to show the six spectral bands, each 2.56 MHz wide, observed by the MWA, and the MWA DS respectively.

\begin{figure*}
    \centering
    \includegraphics[width = 0.44\textwidth]{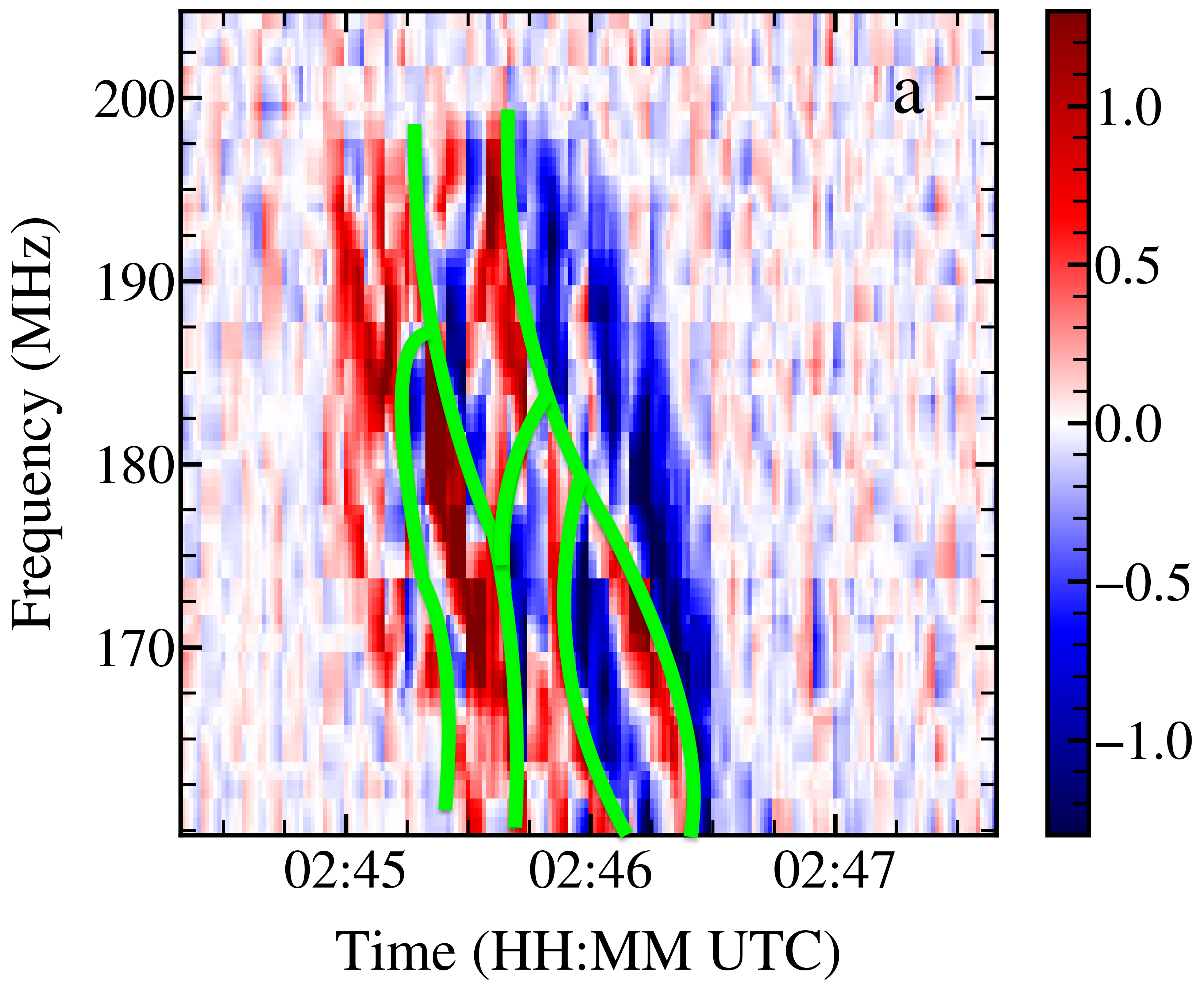}
    \includegraphics[width = 0.44\textwidth]{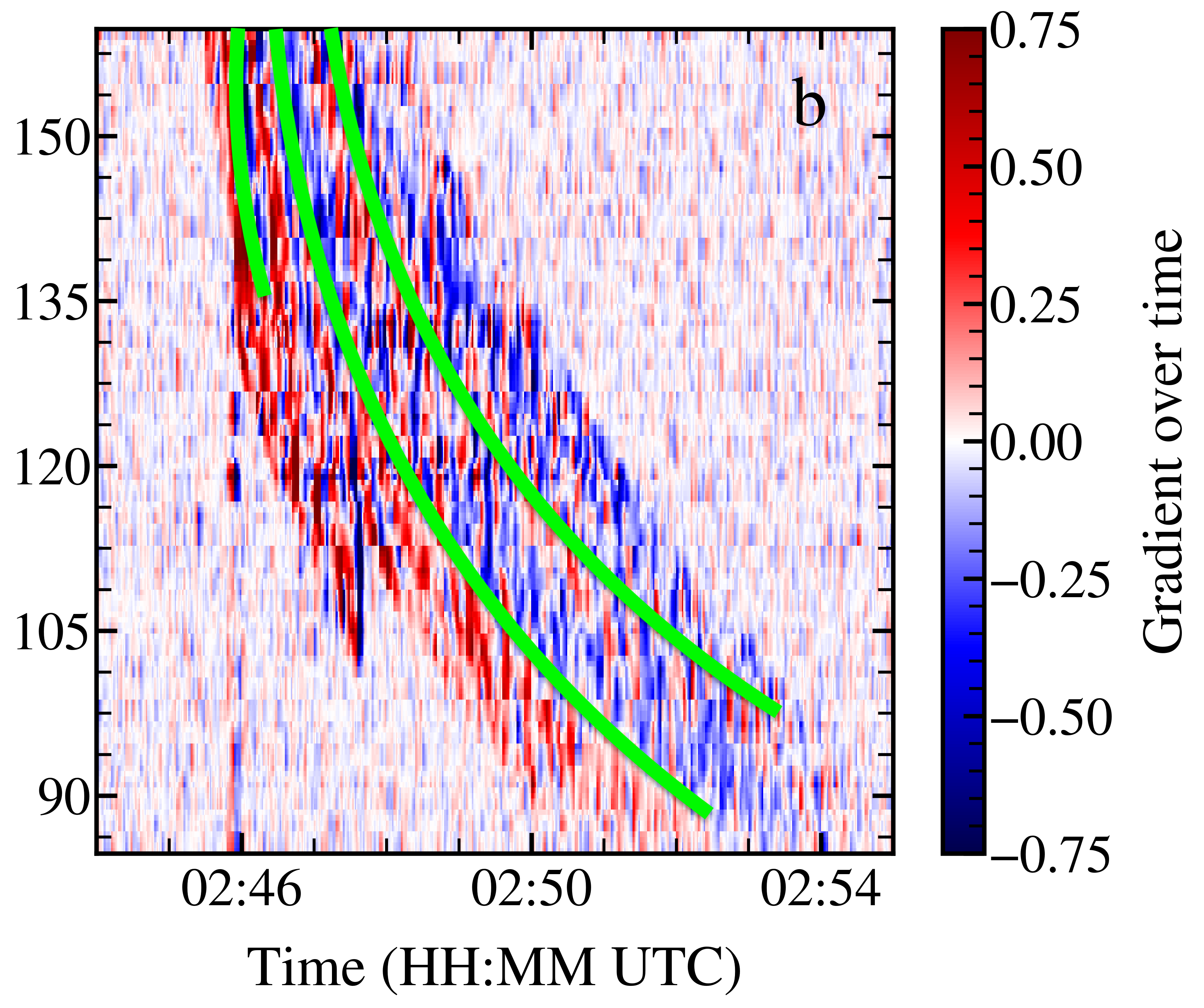}
    \caption{The GraDS of the harmonic emissions, marked by boxes A and B in Figure \ref{fig:hiras_mwa_ds}(a), are shown in the left (a) and right (b) panels, respectively. The green curves mark the positive-to-negative transitions indicating the peaks of emissions. 
    \label{fig:grad_plots}}
\end{figure*}

Identifying morphological similarities between type-II emission lanes can sometimes be difficult due to varying intensities across the lanes and the intrinsic complexity of the emission, in addition to limitations arising from SNR and loss of data due to flagging. Based on the idea that the locus of an emission lane should correspond to the local maximum in the dynamic spectrum, which would, in turn, correspond to a zero crossing in the gradient of the dynamic spectrum, irrespective of the strength of the emission. Hence, in order to systematically identify morphological similarities across lanes, we also examine the dynamic spectrum of the gradient of intensity alongside the usual DS. Gradient across time is taken for individual spectral channels to produce the gradient DS (GraDS). As shown in \ref{fig:grad_plots}, the local maxima, i.e. the transition between the positive (red) to negative (blue) gradients indicate the peak of the emission. Green curves connecting the local maxima have been drawn to make the emission lanes easier to follow. The drift rate and morphology of the harmonic emission changes across the burst. Hence, we divide the emission into two parts, marked by the white dashed boxes in Figure \ref{fig:hiras_mwa_ds}(a), and carry out the gradient analysis separately.    

\section{Results and Comparison with Spectroscopic Imaging using the MWA}\label{sec:res_&_dis}
As shown in Figure \ref{fig:grad_plots}, the GraDS unambiguously marks the transition from the rising (red) to the falling (blue) part of the dynamic spectrum. It systematically identifies the spectro-temporal location of the peaks of the emission lanes even when identifying them directly from the dynamic spectrum is not straightforward. Panels (a) and (b) of Figure \ref{fig:grad_plots} show the GraDSs of the regions marked by the upper (A) and lower (B) white dashed boxes in Figure \ref{fig:hiras_mwa_ds}(a). The locations of these peaks of the emission lanes are marked by green curves in Figure \ref{fig:grad_plots} to guide the eye. Figure \ref{fig:grad_plots}(a) starts with two well-formed parallel lanes, they become non-parallel and branch as the type-II proceeds.
 
The GraDS of the type-II marked box B in Figure \ref{fig:grad_plots}(a) is shown in Figure \ref{fig:grad_plots}(b). Three large-scale positive-to-negative transition regions are identified in the earlier part of the emission. However, in the later part, only two lanes can be identified with similar frequency drifts. It also shows numerous positive-to-negative transitions confined to small spectral and temporal extents, suggesting the presence of multiple narrow-band short-lived fine structures. A comparison of the two panels of Figure \ref{fig:grad_plots} brings out the remarkable difference in the nature of emission between them.

The above exercise shows that this type-II emission does not show band-splitting with morphologically similar emission lanes. Rather it has a complex morphology and for some part has emission lanes that can not be identified clearly. While the harmonic emission marked with box A in Figure \ref{fig:hiras_mwa_ds}(a) shows a clear signature for the presence of multiple lanes, the emission marked by box B in Figure \ref{fig:hiras_mwa_ds}(a) does show two lanes after the first few minutes. The presence of multiple lanes in the earlier part of the emission suggests that the corresponding sources likely lie at different parts of the shock front. The multiple lanes with different frequency drift rates also suggest that the multiple sites of type-II emission are likely moving with different shock speeds. The presence of two bands with similar drift rates after about 02:49 UTC, however, does imply that the emission is coming from regions with similar shock speeds.

\begin{figure}
    \centering
    \includegraphics[scale=0.53]{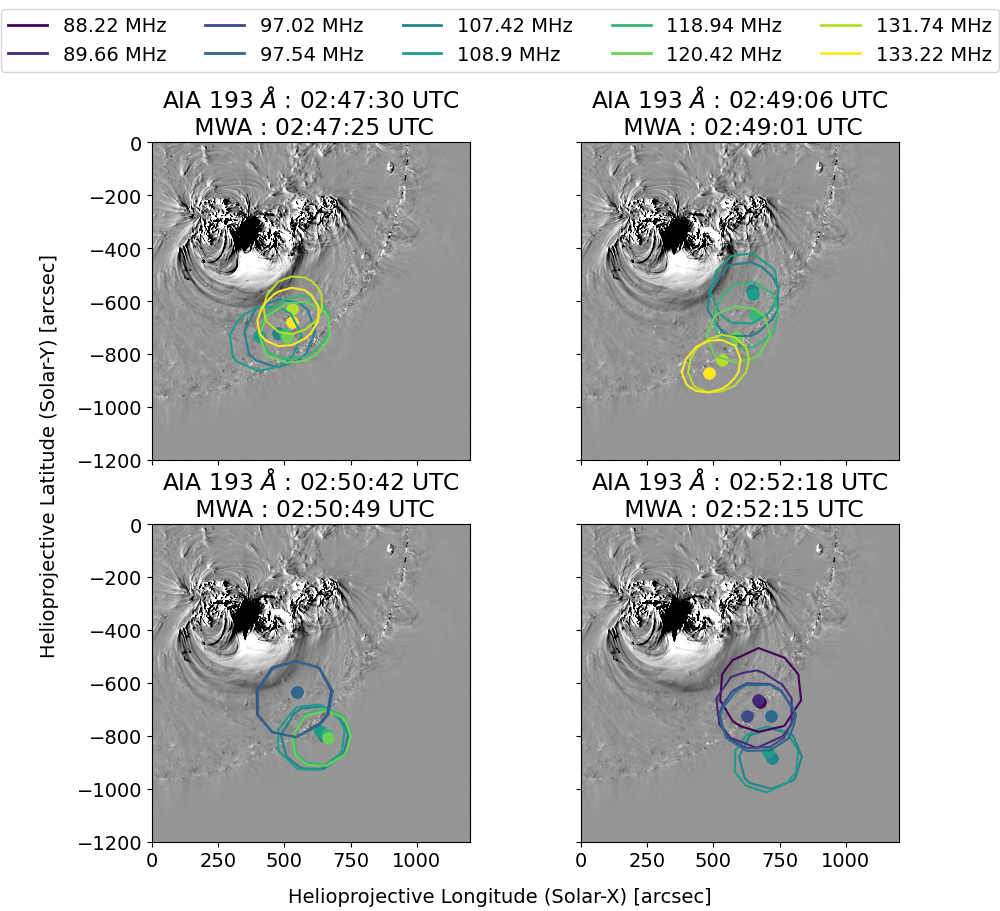}
    \caption{Each of the panels shows locations of the type-II source at different frequencies obtained from MWA spectroscopic snapshot images, overlaid on the AIA base difference images at 193 \r{A} at nearby times. Sources at different radio frequencies are shown in different colors and times corresponding to different panels are marked in Figure \ref{fig:hiras_mwa_ds}(c). The contours and circles correspond to the 90\% of peak intensity and the peak of the best-fit Gaussian used to model the source in MWA radio images. Uncertainties in the location of the peak of the Gaussian source are $\sim$1arcsec and are too small to be visible in this plot.}
    \label{fig:aia_mwa}
\end{figure}

In their study of this type-II, \citet{Bhunia_2023} used the visually estimated similarity of the two lanes to regard this as a case of split-band emission. Their imaging analysis, however, revealed radio sources corresponding to the two lanes to be located far apart and moving in different directions with different speeds. This led them to conclude that this event was inconsistent with expectations from the upstream-downstream scenario. Having started from the hypothesis of band-splitting, \citet{Bhunia_2023} limited themselves to study the location of radio sources corresponding to the peaks of the two split bands. As shown in Figure \ref{fig:grad_plots}, the emission shows complex morphology, and we have just presented evidence for type-II emission to be coming from different parts of the shock. It would, hence, be interesting to investigate the locations of the sources corresponding to the full spectral extent of the emission across time. This exercise can potentially shed light on the type-II source regions as well as their evolution with time. Figure \ref{fig:aia_mwa} shows the locations of the type-II radio sources obtained by spectroscopic snapshot imaging using the MWA. Due to the high signal-to-noise ratio (SNR), the astrometric accuracy of the location of these sources is much smaller than the point-spread-function (PSF) of the instrument. Astrometric accuracy is given as, $\frac{PSF}{2\times SNR}$, which is $\sim1$ arcsecond, and the separation between the sources at different frequencies is much larger than the astrometric accuracy. This clearly shows that the type-II sources at different frequencies are spatially separated and confirms that they could not arise from the upstream-downstream scenario. The locations of these sources do not seem to be random but rather follow well-defined spectral trends, which are currently under investigation.

\section{Conclusion}\label{sec:conclusion}
Type-II solar radio bursts are signatures of shocks. Type-II band-splitting satisfying the upstream-downstream scenario \citep{Vrsnak2001,Vrsnak2002,Vrsnak2004} is a powerful tool for estimating magnetic fields at shock locations. 
However, band-splitting is not commonly observed in type-IIs. Hence, it is important to carefully verify that it is indeed the case for the event under consideration before interpreting it as one. Naturally, incorrectly identifying type-IIs to have split-band emission is likely to lead to incorrect estimates of shock strengths and magnetic fields.

While spectroscopic imaging observations offer the most robust way to unambiguously distinguish between the upstream-downstream and multiple sources on the shock front scenarios, such observations are comparatively rare. In this article, we present a systematic approach to augment the usual study of DS to identify morphological similarities of type-II emission lanes by also examining the GraDSs. Our analysis, based on this approach, found this type-II emission does not follow the upstream-downstream scenario, without relying on imaging and confirms the imaging-based results from \citep{Bhunia_2023}. This confirmation establishes the usefulness of the GraDS-based morphological identification of type-II band-splitting and we hope for wider use of this approach in the community for identifying split-band type-IIs. Our imaging analysis also shows that, for the type-II studied here, the emission at different frequencies comes from a spatially extended region.

\begin{acknowledgements}
We thank the National Institute of Information and Communications Technology (NICT) for providing the HiRAS data publicly. This scientific work makes use of the Murchison Radio-astronomy Observatory (MRO), operated by the Commonwealth Scientific and Industrial Research Organisation (CSIRO). We acknowledge the Wajarri Yamatji people as the traditional owners of the Observatory site.  Support for the operation of the MWA is provided by the Australian Government's National Collaborative Research Infrastructure Strategy (NCRIS), under a contract to Curtin University administered by Astronomy Australia Limited. We acknowledge the Pawsey Supercomputing Centre, which is supported by the Western Australian and Australian Governments. P.M. and D.O. acknowledge the support of the Department of Atomic Energy, Government of India, under the project no. 12-R\&D-TFR-5.02-0700.  D.K. acknowledges the support by the NASA Living with a Star Jack Eddy Postdoctoral Fellowship Program, administered by UCAR’s Cooperative Programs for the Advancement of Earth System Science (CPAESS) under award \#80NSSC22M0097. D.K. thanks the organizers of the IAU Symposium 388 for providing travel support to attend the symposium and present this work on behalf of the first author.    
\end{acknowledgements}

\bibliography{Sample}{}
\bibliographystyle{aasjournal}

\end{document}